\documentclass[letterpaper]{article} 
\usepackage{aaai25}  
\usepackage{times}  
\usepackage{helvet}  
\usepackage{courier}  
\usepackage[hyphens]{url}  
\usepackage{graphicx} 
\usepackage{natbib}  
\usepackage{caption} 
\usepackage{algorithm}
\usepackage{algorithmic}

\usepackage{newfloat}
\usepackage{listings}
\DeclareCaptionStyle{ruled}{labelfont=normalfont,labelsep=colon,strut=off} 
\floatstyle{ruled}
\newfloat{listing}{tb}{lst}{}
\floatname{listing}{Listing}
\usepackage{amsmath}
\usepackage{amssymb}
\usepackage{cancel}

\usepackage{xcolor}
\usepackage{ulem}

\title{Generating Quantum Reservoir State Representations with Random Matrices}
\author{
    Samuel Tovey\equalcontrib\textsuperscript{\rm 1}, 
    Tobias Fellner\equalcontrib\textsuperscript{\rm 1}, 
    Christian Holm\textsuperscript{\rm 1}, 
    Michael Spannowsky\textsuperscript{\rm 2}\\
}
\affiliations{
    \textsuperscript{\rm 1}    
    Institute for Computational Physics\\
    University of Stuttgart\\
    Stuttgart, Germany, 70569 \\
    
    \textsuperscript{\rm 2}     
    Institute for Particle Physics Phenomenology\\
    Department of Physics\\
    Durham University\\
    Durham, DH1 3LE, U.K \\ 
    stovey@icp.uni-stuttgart.de, tfellner@icp.uni-stuttgart.de
}

\begin{document}

\maketitle

\begin{abstract}
We demonstrate a novel approach to reservoir computation measurements using random matrices.
We do so to motivate how atomic-scale devices could be used for real-world computational applications.
Our approach uses random matrices to construct reservoir measurements, introducing a simple, scalable means of generating state representations.
In our studies, two reservoirs, a five-atom Heisenberg spin chain and a five-qubit quantum circuit, perform time series prediction and data interpolation.
The performance of the measurement technique and current limitations are discussed in detail, along with an exploration of the diversity of measurements provided by the random matrices.
In addition, we explore the role of reservoir parameters such as coupling strength and measurement dimension, providing insight into how these learning machines could be automatically tuned for different problems.
This research highlights the use of random matrices to measure simple quantum reservoirs for natural learning devices, and outlines a path forward for improving their performance and experimental realization.
\end{abstract}

\section{Introduction}
Computing with physical systems such as particle swarms~\cite{lymburn21a, wang24a} or lasers~\cite{duport12a, nakajima21a} has emerged as a promising candidate for a new wave of technologies built around better utilizing natural systems as learning devices.
These approaches to computation, broadly captured under the term Extreme Learning Machines (ELMs)~\cite{huang14a}, of which Reservoir Computing (RC) can be seen as a subset, suggest that the majority function of highly parameterized learning devices is simply the projection of information into high dimensions, from there, a much smaller, trainable readout layer may be tuned to make predictions on this projection.

One promising research direction is the use of quantum devices for reservoir computing~\cite{fujii_harnessing_2017, tanaka19a, mujal21a, innocenti23a}. Quantum systems studied include spin-lattices~\cite{cindrak24a}, superconducting qubits~\cite{suzuki22a, dudas23a}, or artificial atoms~\cite{bravo22a}, which are viable candidates for future, real-world devices due to their size and solid-state nature.
A convenient aspect of utilizing quantum devices for this form of computing comes in their power-law scaling of states as a function of particles.
This factor makes relatively small quantum devices well-suited for projecting input signals into high-dimensional spaces, which can then be read off for predictions.
To this end, there have been numerous demonstrations of the efficacy of quantum RC methods ranging from excited state energy predictions in molecules~\cite{domingo22a}, quantum reinforcement learning~\cite{chen24a}, and object or image classification~\cite{suzuki22a, spagnolo22a} to mention just a few.
A critical problem, however, is the construction of the state representations of the learning machine.
This representation must comprise a large set of numbers that provide a complete or over-complete representation of the system.
In the case of quantum systems, the space of well-defined measurements that could be used in this representation is small, and in most cases, only the Pauli spin matrices are used on individual spins~\cite{domingo23a, mujal22a, pena23a} to construct such a state representation.
This approach results in small state representations and, therefore, less expressive measurements.
In this work, we suggest utilizing random measurements on the reservoir as an alternative to the Pauli spin matrices.
This approach is motivated by the fact that the state representation passed into the reservoir's readout layer need not be interpretable; rather, it should be sensitive to changes in the internal state.
Therefore, rather than restricting measurements to the Pauli spin matrices, we explore a fixed set of random measurements.
This can be achieved by constructing random hermitian matrices, which can then act on the systems' state in the presence of a time-varying input.
In this way, large numbers of diverse measurements can be made on the reservoir, producing larger state representations, which, we argue, aids in better reservoir computing.

Incorporating Random Matrix Theory (RMT) into the framework of quantum reservoir computing (QRC) here presents a novel approach to harnessing the intrinsic complexities of quantum systems for computational tasks. RMT, with its roots deeply embedded in statistical physics and mathematical theories, provides a versatile tool for understanding the statistical behaviors of complex systems \cite{Guhr:1997ve,anderson2010introduction,fyodorov2010introduction,Livan_2018}, particularly those governed by quantum mechanics. 
The application of RMT in this context is motivated by the unique ability of random matrices to capture the dynamic properties of quantum systems, thereby enriching the state representations used in reservoir computing.
Using random matrices to generate reservoir state representations in quantum systems can increase the richness of a reservoir's state space can significantly enhance its computational capacity. This strategy not only expands the expressive power of the state representations but also aligns with the core philosophy of reservoir computing, where the non-linearity and high-dimensionality of the reservoir's dynamics are crucial for processing and predicting complex time-series data. 

Previous work has explored how the scaling of the Hilbert space can improve reservoir performance~\cite{kalfus22a} as well as the role of certain kinds of quantum interactions such as entanglement~\cite{goetting23a}.
Some work has looked directly at the measurement problem as in Mujal \textit{et. al.}~\cite{mujal23a} where they study to use of intermittent measurements to produce a more reliable workflow for online RC.
In their 2022 work, Domingo \textit{et. al}~\cite{domingo22a} demonstrated that random quantum circuits could be used to construct state representations from molecular-input data.
Such an approach is similar to the work presented here, albeit requiring a larger number of implemented quantum circuits.
In all cases, these investigations have utilized the standard set of observables for the state representation, typically the Pauli spin matrices.
While this allows for a more interpretable state representation, the important aspect of the representation is its sensitivity to changes with respect to the evolution of the reservoir.
In other recent work by Haug \textit{et. al.}~\cite{haug23a} and Elben \textit{et. al.}~\cite{elben19a}, random measurements were used to describe quantum support vector machines and quantum many-body systems, respectively, to great effect in a similar vein to what is presented here.
This work introduces a more generalizable approach to state measurement wherein random matrices are used as measurements on the reservoir to construct a vector representation.

The remainder of this paper is organized as follows.
First, we introduce the fundamental framework of reservoir computing and present the two reservoir systems studied in this work, a five-atom spin chain and a gate-based quantum circuit. We then discuss the formalism of a quantum reservoir state representation. 
Subsequently, we present the results of our investigations on several test problems, which aim both to demonstrate the capability of the measurement method and to better understand the effects of tuning certain model parameters.
Finally, we discuss the practical limitations and give an outlook on future research directions.

\section{Framework}
\subsection{Reservoir Computing}
\label{subsec:reservoir-computing}
Reservoir computing is an approach to machine learning where the non-linear dynamics of a system are used to embed some input signal into a high dimensional space from which it can be read out by a trainable readout layer~\cite{gauthier21a}.
They were initially proposed based on the so-called echo state networks~\cite{jaeger04a} but have since expanded well beyond their humble origins~\cite{tanaka19a}.
In a typical neural network architecture, hidden layers play the role of a high-dimensional embedding
\begin{equation}
    \centering
    f:\mathcal{A}^{N} \rightarrow \mathcal{B}^{M} : M > N
    \label{eqn:embedding}
\end{equation}
where $A$ and $B$ are mathematical spaces.
After being passed through these layers, data is read out by a simple linear combination
\begin{equation}
    \centering
    y_{i} = w_{ij}x_{j}
\end{equation}
where $y_{i}$ is the $i^{\text{th}}$ output of the neural network and $w_{ij}$ the weight matrix of the final layer.
During training, this embedding is trained on the data such that for each data point passed through the network, a unique representation of this point in the high dimensional space is produced.
In this work, we often found slightly improved results by allowing one additional layer with a ReLU non-linearity, still vastly reducing the size of the networks but increasing the complexity of the reservoir readout.
In the case of the spin chains, a single ReLU activation is used in the readout layer whereas in the quantum circuit reservoir, no non-linearity was required.

In reservoir computing, rather than using a trainable embedding, a fixed mapping (the reservoir) is selected to perform the role of Equation~\ref{eqn:embedding}.
This mapping can be as simple as a large random matrix, however, in recent years, it has been shown that using dynamical systems as a reservoir provides access to a non-linear function in a high-dimensional space.
These reservoirs work by interacting with their inputs, typically by some coupling of their dynamics.
For example, in the case of Swarm computing, a predator particle is driven through a swarm of reservoir particles, which, in turn, avoids the incoming predator~\cite{lymburn21a, wang24a}.
The coupling of these two dynamical systems makes it possible to predict the movement of the predator using some state representation of the swarm.
Beyond the immediate benefit of requiring far less effort in training, reservoir computing also naturally incorporates memory into the prediction process.
Most systems used in reservoir computing display non-markovian dynamics, that is, their movement can be very sensitive to their initial conditions and depend on their states far into the past~\cite{inubushi17a}.
Not only does this result in a non-linear mapping, it also provides a means for predicting time series problems that also display this kind of memory~\cite{fang23a}.

In this study, the first reservoir studied is a five-atom spin chain, simulated in the QuTIP engine~\cite{johannson12a, johannson13a}.
The spin chain is modeled by the time-dependent Quantum Heisenberg Hamiltonian
\begin{equation}
\begin{split}
    \mathcal{H} = &-\frac{1}{2} J \sum\limits_{\langle ij \rangle} (\sigma_x^{(i)} \sigma_x^{(j)} + \sigma_y^{(i)} \sigma_y^{(j)} + \sigma_z^{(i)} \sigma_z^{(j)}) \\
    &- \frac{1}{2} h(t) \sum\limits_{l} \sigma_z^{(l)}\,,
\end{split}
\end{equation}
where $J$ is the spin coupling constant, $\sigma_k^{(m)}$ with $k \in \{x, y, z\}$ are the Pauli spin matrices acting on spin $m$, and $h(t)$ is a time varying magnetic field.
The summation over $\langle ij \rangle$ includes all nearest neighbors in the spin chain without boundary condition, while the summation over $l$ includes all spins.
The driving signal is encoded into the reservoir through the magnetic field as it couples to the spins in the chain.
The unitary time evolution is simulated in QuTIP.

Additionally, we utilize a quantum circuit of five qubits as a quantum reservoir. 
We sequentially encode sequences of data-points from a time series into the quantum state by angle-encoding~\cite{havlivcek2019supervised}. 
In this method, a rotation gate is acting on qubits with an angle corresponding to the value of a data point $x_i \in \mathbb{R}$. In the course of this work, the encoding unitary is the tensor product of Pauli-X rotation gates $R_x$ over $n$ qubits
\begin{equation}
\label{eq:angleencoding}
    U_\mathrm{ent}(x_i) = \bigotimes_{j=1}^n R_x(x_i)\,.
\end{equation}
A further unitary operation on the quantum state is applied in order to realize a versatile state representation. 
The circuit is simulated using the Python library Pennylane~\cite{bergholm2018pennylane}.

\subsection{State Representation}
\label{subsec:state-description}
Reservoir computing relies on the ability to embed information from an input into a system that distributes it across many degrees of freedom.
This is, however, not difficult to achieve and therefore, not the bottleneck in reservoir computing research.
Rather, developing a state representation to measure this distribution of information from the system is a far more tedious task, particularly in the field of quantum reservoir computing, where the list of known observables can be painfully small.
In this work, we set out to discuss whether random matrices can be used for reservoir computing on very small quantum systems. 
Therefore, we do not address these limitations in great detail but instead, provide an outlook at the end of the paper about how they could be overcome or even used in future investigations.

In our approach, a set of random hermitian matrices is used to make measurements of a quantum system evolving under a driven Hamiltonian.
Consider the indexed set of random matrices
\begin{equation}
    \centering
    \mathcal{M} = \{ \mathcal{O}_{i} : \mathcal{O}^{\dagger}_{i} = \mathcal{O}_{i} \hspace{0.1cm}\forall\hspace{0.1cm} 0 \leq i \leq N \text{ and } i \in \mathcal{Z} \},
\end{equation}
referred to hereafter as the \textit{measurement set}.
The state representation, $s \in \mathcal{R}^{N}$ can then be constructed by acting on a quantum system using the measurement set with
\begin{equation}
    \centering
    s_{i} = tr\left( \rho \mathcal{O}_{i} \right),
    \label{eqn:measure}
\end{equation}
where $\rho$ is the density matrix of the system.
In a real quantum reservoir, the density matrix $\rho$ is a product over subsystems
\begin{equation}
    \centering 
    \rho = \bigotimes_{m}\rho_{m},
\end{equation}
where $\rho_{m}$ is the density matrix of the $m^{\text{th}}$ subsystem.
Measurements like those in Equation~\ref{eqn:measure} would destroy the system state under study, thereby damaging the reservoir.
Furthermore, larger measurements of the reservoir can be more expensive, particularly on a computer. 
Therefore, it may be favourable to reduce this complexity and measure local states.
In practice, this is represented by substituting the trace operation for the partial trace
\begin{equation}
    \centering
    m^{j}_{i} = tr\left( \rho_{m} \mathcal{O}_{i} \right),
\end{equation}
where $\rho_{m} = tr_{\rho_{k}}\rho$ and $\rho_{k} = \bigotimes_{l}\rho_{l} : l \neq m$ is understood to be the composite formed without the test subsystem.
With this approach, each test subsystem will be assigned its own state representation, $s^{\alpha} \in \mathcal{R}^{P_{\alpha}\leq N}$.
The reasoning is that one can now decide whether or not to apply all observable measurements to all test subsystems or choose some distribution strategy.
The combined state representation, $g$, will have the dimension $F = \bigoplus_{\alpha}P_{\alpha} : F \geq N$.
Combining these state representations into a readout is then handled by the readout, or neural, layer of the reservoir.
In this work, we randomly select test sites for each observable, $\mathcal{O}_{i}$, and concatenate their expectations before passing them to the readout layer.

\section{Experiments}
\subsection{Observable Study}
In the studies presented here, we employ either coupled spins or qubits within a quantum circuit.
In such composite systems, the possible states attainable scales by the power of the number of elements, i.e, for a combined state, $|\psi\rangle$
\begin{equation}
    |\psi\rangle = \bigotimes\limits_{i}^{N}|\phi\rangle_{i} \in \mathcal{H}^{2^{N}},
\end{equation}
for $N$ spins and Hilbert space, $\mathcal{H}$.
When computing measurements in this space, one can apply a sufficiently large matrix and compute a value for the full state. 
However, measurements over subsets are also possible, improving computation time and reducing the problems associated with wavefunction collapse.
As an observable's outcome is related to its eigenspectrum, it is relevant to identify the impact of partial measurements on constructing a state representation and whether performing partial trace measurements improves or degrades the representation.
To do so, we perform two experiments.
In the first, observable matrices of different sizes are randomly produced and their spectra compared, and in the second, these random matrices are used to compute observables on random density matrices with and without partial traces.
The maximum size matrix is fixed by the theoretical maximum number of coupled spin sites considered during our investigations, which was 9.
Therefore, in the largest case, a 512x512 matrix is considered.
All matrices are constructed using the QuTIP \texttt{random\_herm} and \texttt{random\_dm} functions.
\begin{figure}
    \centering
    \includegraphics[width=\linewidth]{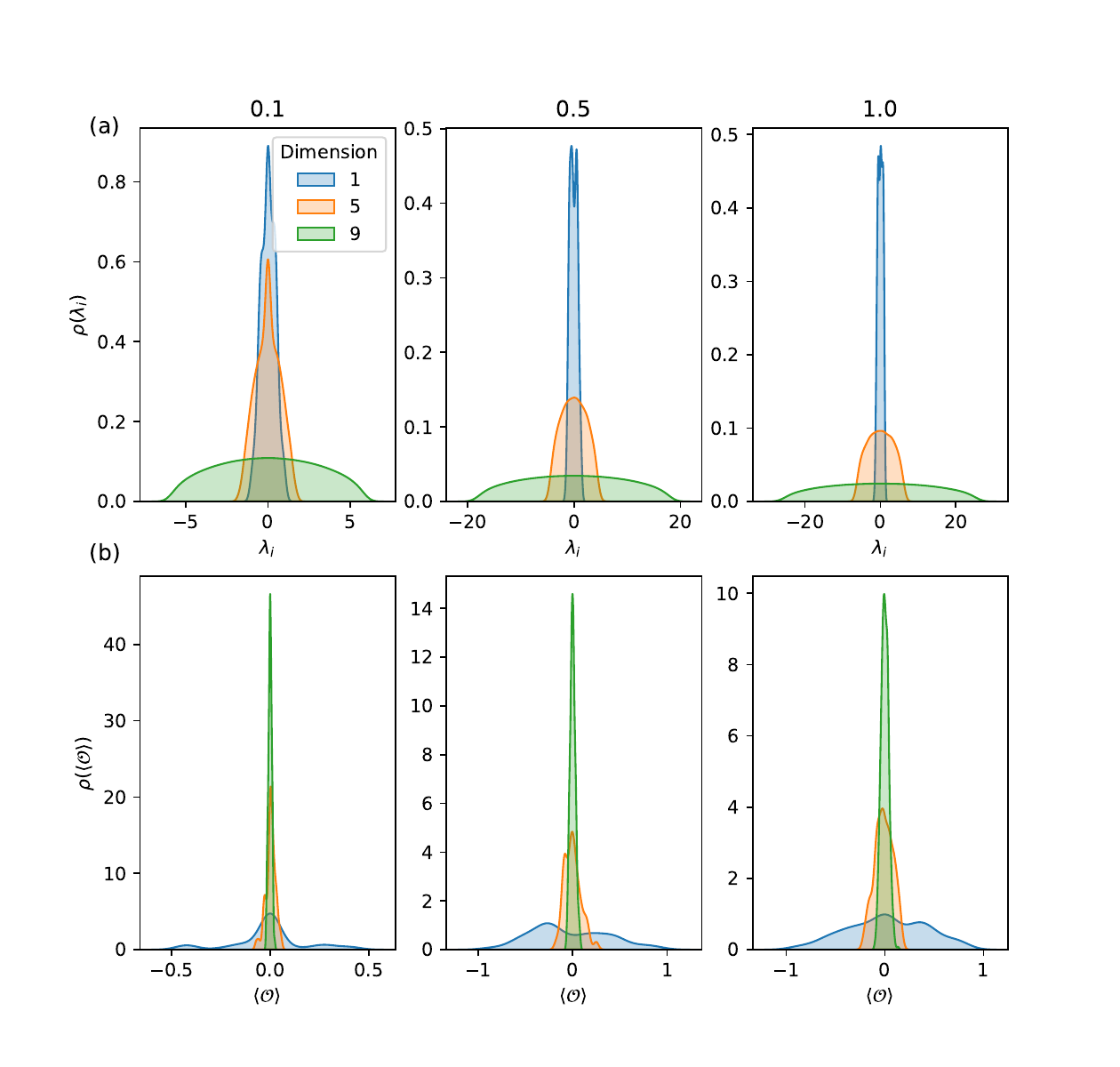}
    \caption{(a) Distribution of random matrix eigenvalues used for the observables computations during the QRC studies. (b) Distribution of the expectation values of observables computed using random matrices of varying sizes and densities. For matrices smaller than 512, measurement sites are chosen randomly.}
    \label{fig:diversity}
\end{figure}
\paragraph{Matrix Statistics}
Figure~\ref{fig:diversity} (a) outlines the results of the first investigation.
In these figures, each frame shows the results of the computation on a set of different dimensional observables constructed with a fixed density, stated in the title of the frame.
For a low density, the random matrix is mostly populated with zeros.
This figure shows that the larger matrices, perhaps unsurprisingly, produce a more diverse range of eigenvalues.
This result suggests that the space of solutions for these larger matrices is larger and, therefore, more ideal for measurements.
However, in real studies, the observable is applied to a state, and therefore, it is important to understand what effect this will have on the measurement's outcome.
\paragraph{Measurement Statistics}
The next component to study is the outcome of measurements on randomly generated states.
This study is performed by fixing a set of measure sites in a Hilbert space built from nine individual spin states, producing a net dimension of 512.
A set of random observables is then computed on these sites via the partial trace.
This is repeated for new sites and observables of larger dimensions.
Results of this study are displayed in Figure~\ref{fig:diversity} (b).
In this case, we see that the smaller observables, computed on partial traces of the full systems, produce more diverse measurements than the larger ones.
This could be due to the states' self-averaging, but it is important for how we construct the state representations in the following studies.
Specifically, these results suggest that we should use the computationally inexpensive option of many measurements of partial traces of the reservoir.
\subsection{Cosine Wave}
As a simple investigation, we train the spin chain to fit a cosine wave and vary the properties of the reservoir to identify the effect on performance.
The initial aim is to show whether the reservoir, measured using random matrices, can accurately reproduce the input function.
To produce the data, a cosine wave is fed into the five-atom reservoir through the magnetic field, coupling to all spins in the chain.
After the simulation, a set of random measurements is made to construct a state representation for each time step, which may then be fed into a readout layer for future predictions.
\begin{figure}
    \centering
    \includegraphics[width=\linewidth]{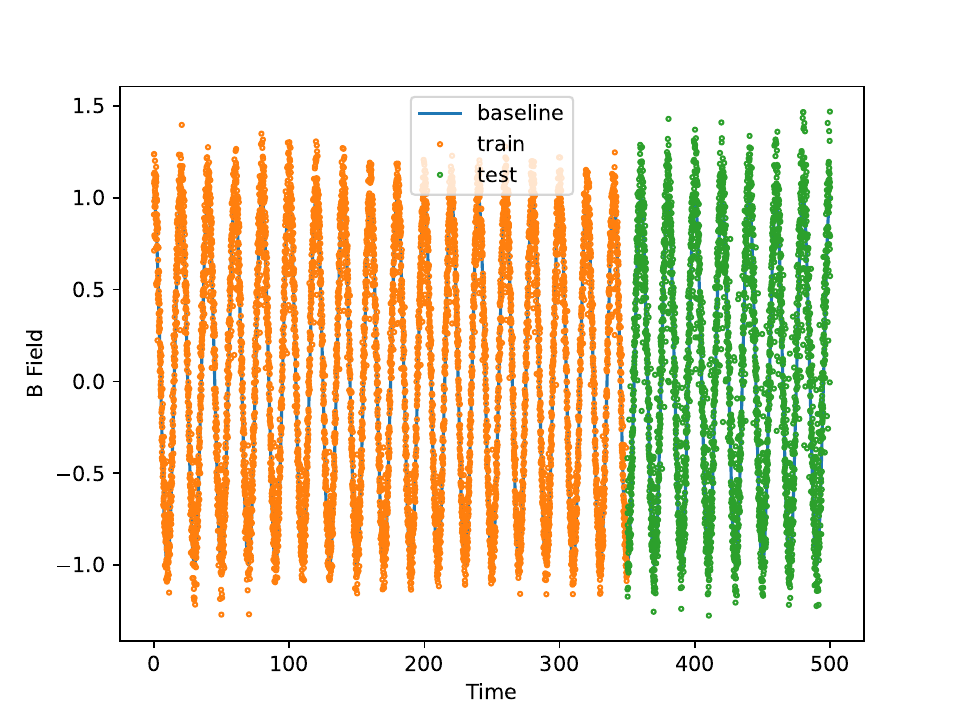}
    \caption{Reservoir predictions of the sine wave. The orange data corresponds to what was passed into the reservoir during training, whereas the green line shows the predictive-only regime.}
    \label{fig:sine-pred}
\end{figure}
\begin{figure*}
    \centering
    \includegraphics[width=\linewidth]{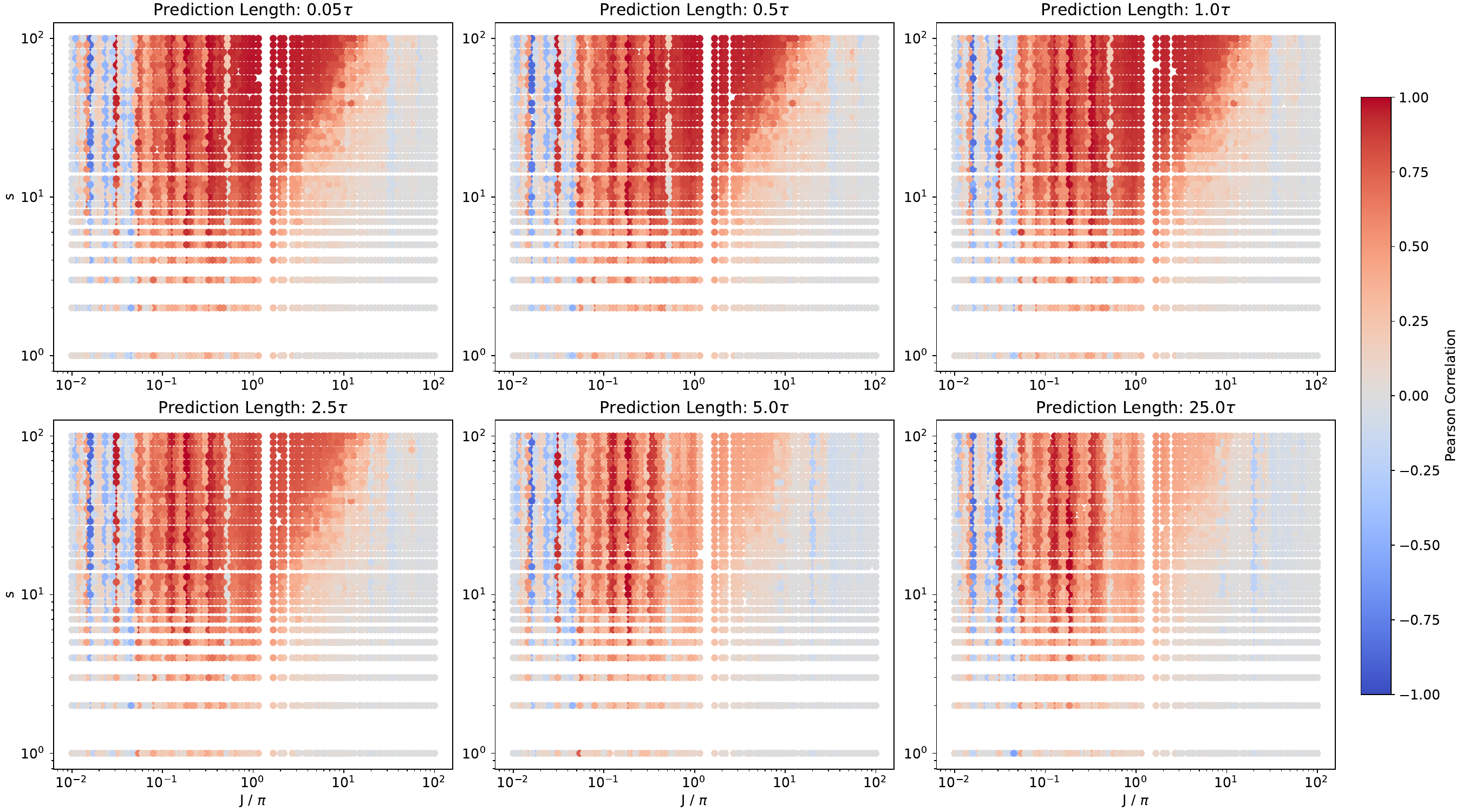}
    \caption{Scan over reservoir parameters for the prediction of the Cosine wave. $\tau$ is a period of the cosine wave and the color corresponds to the Pearson correlation coefficient between the reservoir predictions and the ground truth data. Red indicates a better correlation, whereas blue indicates anti-correlation.}
    \label{fig:sine-wave}
\end{figure*}
Figure~\ref{fig:sine-pred} shows the results of this study for a strongly coupled ($10\cdot \pi$) spin chain on single-step, open-loop prediction.
In open-loop prediction, the real underlying cosine data is still fed into the reservoir, but the readout layer predicts future steps at previously unseen times.
While the results look reasonable, we found that adjusting parameters such as the coupling strength between the spins and the dimensions of the state representation could significantly impact reservoir performance.
To illuminate this effect, we studied how the coupling strength, state representation, and prediction length relate to the system's overall performance.
To do so, we ran a scan of these parameters with coupling strength $J \in \{0.01, 0.1, 0.5, 1.0, 2.0, 5.0, 10.0, 50, 100.0\}$, state dimension $s \in \{1, 10, 20, 50, 100, 500\}$ and prediction length $\tau \in \{0.05\tau, 0.5\tau, 1.0v, 2.5\tau, 5.0\tau, 25.0\tau\}$where $\tau$ is the period of the driving wave.
Figure~\ref{fig:sine-wave} outlines the results of this investigation.
Looking first at the prediction length, we see that the performance decreases as the reservoir must predict further into the future. 
This is expected. 
However, it is interesting to note that the model can make relatively stable predictions even at 2.5 times the period.
The relationship between the reservoir parameters is perhaps of more interest.
We see a stable region of coupling strengths and state dimensions within which the reservoir can produce well-correlated predictions to the underlying signal.
This region is relatively unaffected by prediction length.
Interestingly, there appears to be a slight tradeoff between coupling strength and state dimension for the higher spin couplings.
As this exists for all prediction lengths, it does appear that a larger state dimension may be required to map the more strongly correlated reservoirs correctly.
It should be noted that these coupling strengths do not explore any strongly correlated regime, as they respond only to mild driving signals.
This is largely due to limitations on the simulation time but is an avenue that should be explored in future work.
\subsection{Stock Data Interpolation}
\begin{figure*}
    \centering
    \includegraphics[width=\linewidth]{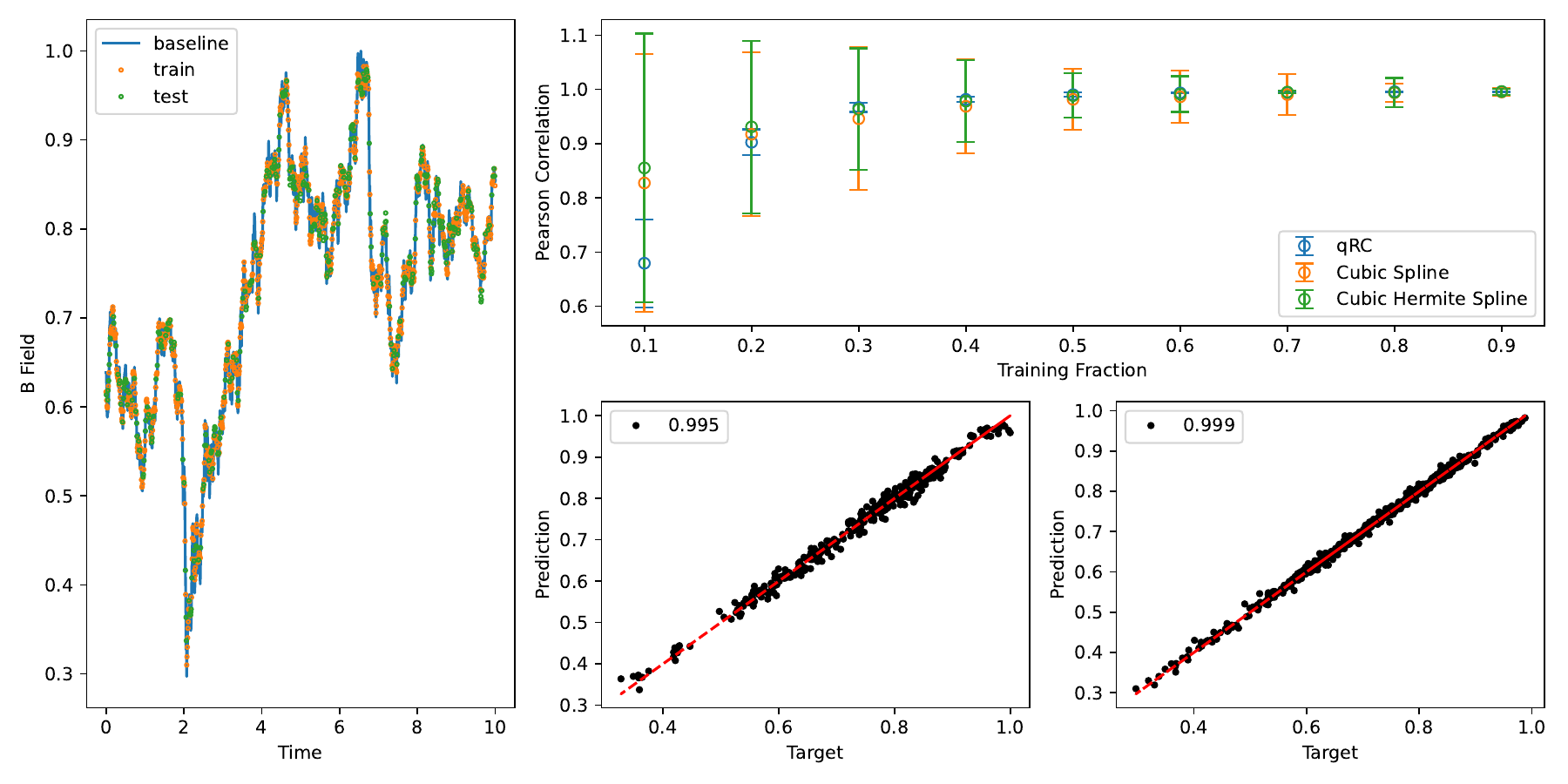}
    \caption{(left) Interpolation of the Dow30 price using the QRC and a 70 \% split. (top right) Pearson correlation coefficients for QRC, Cubic spline, and Cubic Hermite spline methods as a function of the train fraction. (bottom right) Sample Correlation diagrams from the QRC fitting on 70 \% train splits for test (left) and train (right) data. }
    \label{fig:dow30}
\end{figure*}

The subsequent study we performed using the spin chain was price interpolation on the Dow Jones Industrial Average stock market index (Dow30) of daily opening prices.
We do so to demonstrate the ability of the reservoir to perform accurate data interpolation on stochastic time-series data.
The data is fed into a five-atom spin chain through the magnetic field with a $2\pi$ coupling strength.
We then use a 100-dimensional state representation to predict stock prices one-time step ahead (one day in real-time), albeit on shuffled data.
This means the reservoir is used purely as an interpolation device, not for extrapolation.
To study the performance, we train the readout layer of the reservoir on varying fractions of the overall dataset, $x \in \{0.1 .. 0.9\}$, and compute the Pearson correlation coefficient on the test data each time.
This compares with two prominent interpolation methods, Cubic Splines and Cubic Hermite Splines, to judge the efficacy.
The results are displayed in Figure~\ref{fig:dow30}.
Interestingly, the QRC method can perform excellent interpolation on as little as 20 \% of the data.
This is comparable to the other spline methods, although these methods do not use a high-dimensional representation of the reservoir but rather simple 1-D interpolation.
We also attempted to perform direct future prediction on the input data; however, the approach appears to be much more capable of interpolation than extrapolation. 
This also suggests that some memory elements are poorly characterized in the current implementation and could be improved in future work.
\subsection{Quantum Circuit Reservoir Computing on the Mackey-Glass Series}
\begin{figure*}[ht]
    \centering
    \includegraphics[width=\linewidth]{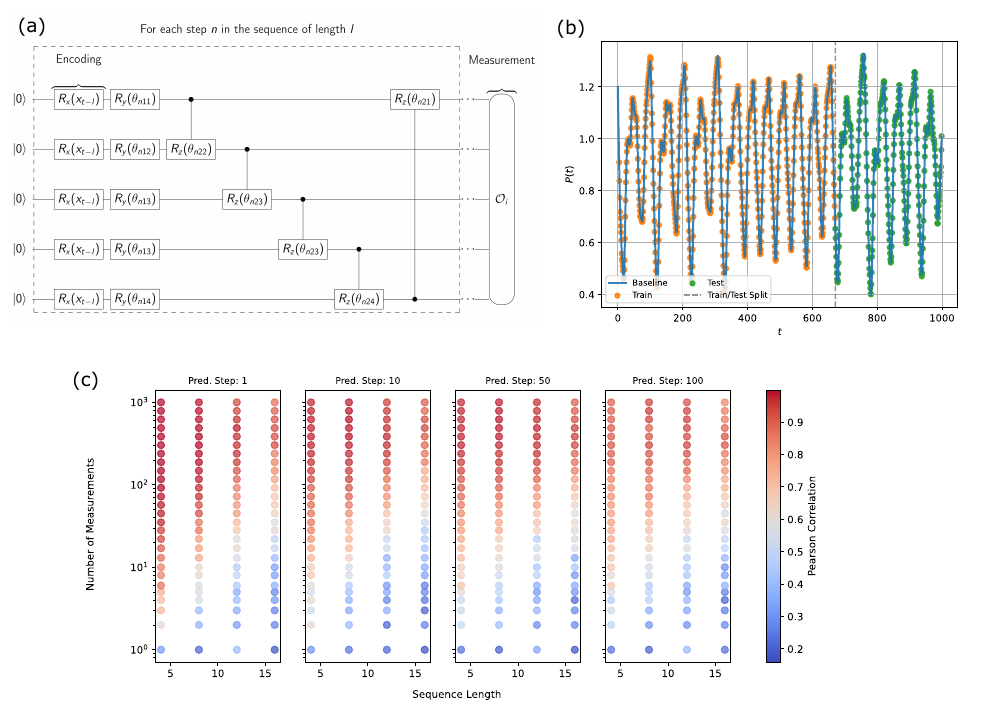}
    \caption{
        (a) Quantum circuit used in this study. A sequence of data points is sequentially encoded into the quantum state of the five-qubit system. Single and multi-qubit gates with fixed rotation parameters $\theta_i$ act on the quantum state. At the end of the circuit, the expectation values of random Hermitian matrices are measured and fed into a linear layer. 
        (b) Mackey-glass Function and the predictions of the trained model with the highest correlation score on the test data. 
        (c) Result of the parameter scan for different sequence lengths, different prediction lengths, and different numbers of random hermitian measurements. Shown is the correlation on the test data for the trained models.
    }
    \label{fig:mackay_qc}
\end{figure*}

For our final investigation, we examine a quantum circuit comprising 5 qubits as a reservoir. 
We train the reservoir to predict the future dynamics of the Mackey-Glass time series $P(t)$~\cite{mackey77a}, which is described by the one dimensional delayed differential equation
\begin{equation}
        \frac{\mathrm{d} P(t)}{\mathrm{d}t} = \frac{\beta P(t-\tau)}{1 + P(t-\tau)^n} - \gamma P(t)\,.
\end{equation}
For the standard parameters $\{\beta, \gamma, \tau, n\} = \{0.2, 0.1, 17, 10\}$ and $P(0)=1.2$ the system shows chaotic behavior.
The time series is generated using a fourth-order Runge-Kutta integrator~\cite{runge95a, kutta01a} at a $1 s$ time step.
The data is split into sequences of length $l$ and sequentially encoded into a state representation by the circuit depicted in Figure~\ref{fig:mackay_qc}~(a). A single data point $x_i$ is encoded via angle encoding as described in Equation~\ref{eq:angleencoding}. In an alternating manner, the encoding and an ansatz of Pauli-Y rotation gates and controlled Pauli-Z rotation gates is executed. The role of the ansatz is to realize a versatile state representation of the input sequence.
The parameters $\theta_{ijk}$ are sampled randomly from a uniform distribution over $[0, 2\pi]$ and are fixed in the subsequent process. 
Following the circuit's execution, a batch of random Hermitian matrices' expectation values is measured. These results are then fed into a classical linear layer, which is trained to predict a future data point.
We conducted a parameter scan for different sequence lengths $l=\{4, 8, 12, 16\}$, different prediction steps $\{1, 10, 50, 100\}$ and different numbers of measurements equally spaced on a logarithmic scale in the interval $[1, 1000]$.
We split the data set into 67\% training and 33\% test data and compute the Pearson Correlation between the actual and predicted test data of the trained reservoir. 
The best result of the parameter scan is shown in Figure~\ref{fig:mackay_qc}~(b). 
The values on the test set are predicted by feeding a sequence of the test data into the reservoir. 
The model achieves a correlation of up to $0.999$.
Figure~\ref{fig:mackay_qc}~(c) displays the results of the parameter scan. 
It's noticeable that with an increased number of measurements, the model attains a higher correlation on the test data. 
Two effects can contribute to this effect. First, the size of the classical output layer increases and the model may gain more flexibility in learning the data.
Second, the model obtains access to a more comprehensive representation of the state representation. 
Moreover, it becomes apparent that by extending the sequence length, the model's performance appears to diminish for a comparable number of measurements. This suggests the necessity for a more expressive model architecture to effectively capture longer sequences.
Lastly, we notice that even for large prediction lengths of 50 or 100 steps, the model can perform predictions on the non-linear time series with a high accuracy. 

\section{Conclusion and Outlook}
We presented a new approach to measuring quantum reservoir computers using random matrices.
Our method is scalable to arbitrary state dimensions and performs well in test cases using two quantum reservoirs, a Heisenberg spin chain, and a quantum circuit.
While the results shown here correspond to quantum systems, any reservoir computing paradigm could be substituted as long as the measurement problem can be formulated in terms of matrix operations.
As a first investigation, we showed the spectrum and observable range of random matrices on random system states. 
We found that partial traces over the entire system may improve the diversity of observables that can be used in reservoir computation.
We then demonstrated the effectiveness of our method on the problem of time series prediction for a sine wave and the Mackey-Glass time series, as well as data interpolation on historical Dow30 data. 
Our results suggest that random matrices work well for different reservoirs for state representation.
However, its practical application to real hardware remains an open question. Currently, it is not feasible to fulfill expectation values over hundreds of measurements with quantum devices. However, there are ways to overcome this challenge. In many real devices, scientists perform measurements far away from the driving mechanism so that multiple measurements can be made without perturbing the primary dynamics. In these cases, the drive signal is often repeated to obtain the required statistics, resulting in a non-real-time workflow. Looking to the future, just as we use chips that are limited by a bit architecture, it's intriguing to imagine that this could be extended to reservoir computing, where small devices such as stacks of spin lattices are designed to perform hundreds of measurements over thousands of ensembles simultaneously.
One avenue for future research is to investigate much more highly correlated spin systems or alternative reservoirs with better learning capabilities. 
In addition, larger reservoirs should be explored to better understand the influence of correlations spanning more degrees of freedom, possibly also in experiments.
Another key element to be studied is the influence of quantum phenomena such as entanglement and the phase space used on the performance of the storage. The aim is to determine whether true quantum phenomena influence reservoir performance. While for a certain QRC architecture an influence of quantum properties on the reservoir performance has been observed~\cite{goetting23a}, these results have to be investigated for the architectures of this paper in future work.

\section{Acknowledgments}
C.H, S.T and T.F. acknowledge financial support from the German Funding Agency (Deutsche Forschungsgemeinschaft DFG) under Germany’s Excellence Strategy EXC 2075-390740016.
C.H, S.T and T.F. acknowledge financial support from the German Funding Agency (Deutsche Forschungsgemeinschaft DFG) under the Priority Program SPP 2363, “Utilization and Development of Machine Learning for Molecular Applications - Molecular Machine Learning” Project No. 497249646.
The authors acknowledge support by the state of Baden-Württemberg through bwHPC and the German Research Foundation (DFG) through grant INST 35/1597-1 FUGG.
The authors acknowledge support from the Deutsche Forschungsgemeinschaft (DFG, German Research Foundation) Compute Cluster grant no. 492175459.

\section{Data availability statement}
The files to reproduce the findings of this study are openly available in the data repository of the University of Stuttgart (DaRUS) in the Dataset "Replication Scripts for: Generating Quantum Reservoir State Representations with Random Matrices": \url{https://doi.org/10.18419/DARUS-4742}.

\bibliography{aaai25}

\begin{thebibliography}{39}
\providecommand{\natexlab}[1]{#1}

\bibitem[{Anderson, Guionnet, and Zeitouni(2010)}]{anderson2010introduction}
Anderson, G.; Guionnet, A.; and Zeitouni, O. 2010.
\newblock \emph{An Introduction to Random Matrices}.
\newblock Cambridge Studies in Advanced Mathematics. Cambridge University Press.
\newblock ISBN 9780521194525.

\bibitem[{Bergholm et~al.(2018)Bergholm, Izaac, Schuld, Gogolin, Ahmed, Ajith, Alam, Alonso-Linaje, AkashNarayanan, Asadi et~al.}]{bergholm2018pennylane}
Bergholm, V.; Izaac, J.; Schuld, M.; Gogolin, C.; Ahmed, S.; Ajith, V.; Alam, M.~S.; Alonso-Linaje, G.; AkashNarayanan, B.; Asadi, A.; et~al. 2018.
\newblock Pennylane: Automatic differentiation of hybrid quantum-classical computations.
\newblock \emph{arXiv preprint arXiv:1811.04968}.

\bibitem[{Bravo et~al.(2022)Bravo, Najafi, Gao, and Yelin}]{bravo22a}
Bravo, R.~A.; Najafi, K.; Gao, X.; and Yelin, S.~F. 2022.
\newblock Quantum Reservoir Computing Using Arrays of Rydberg Atoms.
\newblock \emph{PRX Quantum}, 3: 030325.

\bibitem[{Chen(2024)}]{chen24a}
Chen, S. Y.-C. 2024.
\newblock Efficient Quantum Recurrent Reinforcement Learning Via Quantum Reservoir Computing.
\newblock In \emph{ICASSP 2024 - 2024 IEEE International Conference on Acoustics, Speech and Signal Processing (ICASSP)}, 13186--13190.

\bibitem[{Domingo, Carlo, and Borondo(2022)}]{domingo22a}
Domingo, L.; Carlo, G.; and Borondo, F. 2022.
\newblock Optimal quantum reservoir computing for the noisy intermediate-scale quantum era.
\newblock \emph{Phys. Rev. E}, 106: L043301.

\bibitem[{Domingo, Carlo, and Borondo(2023)}]{domingo23a}
Domingo, L.; Carlo, G.; and Borondo, F. 2023.
\newblock Taking advantage of noise in quantum reservoir computing.
\newblock \emph{Scientific Reports}, 13(1): 8790.

\bibitem[{Dudas et~al.(2023)Dudas, Carles, Plouet, Mizrahi, Grollier, and Markovi{\'{c}}}]{dudas23a}
Dudas, J.; Carles, B.; Plouet, E.; Mizrahi, F.~A.; Grollier, J.; and Markovi{\'{c}}, D. 2023.
\newblock Quantum reservoir computing implementation on coherently coupled quantum oscillators.
\newblock \emph{npj Quantum Information}, 9(1): 64.

\bibitem[{Duport et~al.(2012)Duport, Schneider, Smerieri, Haelterman, and Massar}]{duport12a}
Duport, F.; Schneider, B.; Smerieri, A.; Haelterman, M.; and Massar, S. 2012.
\newblock All-optical reservoir computing.
\newblock \emph{Opt. Express}, 20(20): 22783--22795.

\bibitem[{Elben et~al.(2019)Elben, Vermersch, Roos, and Zoller}]{elben19a}
Elben, A.; Vermersch, B.; Roos, C.~F.; and Zoller, P. 2019.
\newblock Statistical correlations between locally randomized measurements: A toolbox for probing entanglement in many-body quantum states.
\newblock \emph{Phys. Rev. A}, 99: 052323.

\bibitem[{Fang et~al.(2023)Fang, Lu, Gao, and Duan}]{fang23a}
Fang, C.; Lu, Y.; Gao, T.; and Duan, J. 2023.
\newblock Reservoir computing with error correction: Long-term behaviors of stochastic dynamical systems.
\newblock \emph{Physica D: Nonlinear Phenomena}, 456: 133919.

\bibitem[{Fujii and Nakajima(2017)}]{fujii_harnessing_2017}
Fujii, K.; and Nakajima, K. 2017.
\newblock Harnessing Disordered-Ensemble Quantum Dynamics for Machine Learning.
\newblock \emph{Phys. Rev. A}, 8(2): 024030.
\newblock Publisher: American Physical Society.

\bibitem[{Fyodorov(2010)}]{fyodorov2010introduction}
Fyodorov, Y.~V. 2010.
\newblock Introduction to the Random Matrix Theory: Gaussian Unitary Ensemble and Beyond.
\newblock arXiv:math-ph/0412017.

\bibitem[{Gauthier et~al.(2021)Gauthier, Bollt, Griffith, and Barbosa}]{gauthier21a}
Gauthier, D.~J.; Bollt, E.; Griffith, A.; and Barbosa, W. A.~S. 2021.
\newblock Next generation reservoir computing.
\newblock \emph{Nature Communications}, 12(1): 5564.

\bibitem[{G\"otting, Lohof, and Gies(2023)}]{goetting23a}
G\"otting, N.; Lohof, F.; and Gies, C. 2023.
\newblock Exploring quantumness in quantum reservoir computing.
\newblock \emph{Phys. Rev. A}, 108: 052427.

\bibitem[{Guhr, Muller-Groeling, and Weidenmuller(1998)}]{Guhr:1997ve}
Guhr, T.; Muller-Groeling, A.; and Weidenmuller, H.~A. 1998.
\newblock {Random matrix theories in quantum physics: Common concepts}.
\newblock \emph{Phys. Rept.}, 299: 189--425.

\bibitem[{Haug, Self, and Kim(2023)}]{haug23a}
Haug, T.; Self, C.~N.; and Kim, M.~S. 2023.
\newblock Quantum machine learning of large datasets using randomized measurements.
\newblock \emph{Machine Learning: Science and Technology}, 4(1): 015005.

\bibitem[{Havl{\'\i}{\v{c}}ek et~al.(2019)Havl{\'\i}{\v{c}}ek, C{\'o}rcoles, Temme, Harrow, Kandala, Chow, and Gambetta}]{havlivcek2019supervised}
Havl{\'\i}{\v{c}}ek, V.; C{\'o}rcoles, A.~D.; Temme, K.; Harrow, A.~W.; Kandala, A.; Chow, J.~M.; and Gambetta, J.~M. 2019.
\newblock Supervised learning with quantum-enhanced feature spaces.
\newblock \emph{Nature}, 567(7747): 209--212.

\bibitem[{Huang(2014)}]{huang14a}
Huang, G.-B. 2014.
\newblock An Insight into Extreme Learning Machines: Random Neurons, Random Features and Kernels.
\newblock \emph{Cognitive Computation}, 6(3): 376--390.

\bibitem[{\ifmmode~\check{C}\else \v{C}\fi{}indrak et~al.(2024)\ifmmode~\check{C}\else \v{C}\fi{}indrak, Donvil, L\"udge, and Jaurigue}]{cindrak24a}
\ifmmode~\check{C}\else \v{C}\fi{}indrak, S.; Donvil, B.; L\"udge, K.; and Jaurigue, L. 2024.
\newblock Enhancing the performance of quantum reservoir computing and solving the time-complexity problem by artificial memory restriction.
\newblock \emph{Phys. Rev. Res.}, 6: 013051.

\bibitem[{Innocenti et~al.(2023)Innocenti, Lorenzo, Palmisano, Ferraro, Paternostro, and Palma}]{innocenti23a}
Innocenti, L.; Lorenzo, S.; Palmisano, I.; Ferraro, A.; Paternostro, M.; and Palma, G.~M. 2023.
\newblock Potential and limitations of quantum extreme learning machines.
\newblock \emph{Communications Physics}, 6(1): 118.

\bibitem[{Inubushi and Yoshimura(2017)}]{inubushi17a}
Inubushi, M.; and Yoshimura, K. 2017.
\newblock Reservoir Computing Beyond Memory-Nonlinearity Trade-off.
\newblock \emph{Scientific Reports}, 7(1): 10199.

\bibitem[{Jaeger and Haas(2004)}]{jaeger04a}
Jaeger, H.; and Haas, H. 2004.
\newblock Harnessing Nonlinearity: Predicting Chaotic Systems and Saving Energy in Wireless Communication.
\newblock \emph{Science}, 304(5667): 78--80.

\bibitem[{Johansson, Nation, and Nori(2012)}]{johannson12a}
Johansson, J.; Nation, P.; and Nori, F. 2012.
\newblock QuTiP: An open-source Python framework for the dynamics of open quantum systems.
\newblock \emph{Computer Physics Communications}, 183(8): 1760--1772.

\bibitem[{Johansson, Nation, and Nori(2013)}]{johannson13a}
Johansson, J.; Nation, P.; and Nori, F. 2013.
\newblock QuTiP 2: A Python framework for the dynamics of open quantum systems.
\newblock \emph{Computer Physics Communications}, 184(4): 1234--1240.

\bibitem[{Kalfus et~al.(2022)Kalfus, Ribeill, Rowlands, Krovi, Ohki, and Govia}]{kalfus22a}
Kalfus, W.~D.; Ribeill, G.~J.; Rowlands, G.~E.; Krovi, H.~K.; Ohki, T.~A.; and Govia, L. C.~G. 2022.
\newblock Hilbert space as a computational resource in reservoir computing.
\newblock \emph{Phys. Rev. Res.}, 4: 033007.

\bibitem[{Kutta(1901)}]{kutta01a}
Kutta, W. 1901.
\newblock Beitrag zur n\"aherungsweisen {I}ntegration totaler {D}ifferentialgleichungen.
\newblock \emph{Zeit. Math. Phys.}, 46: 435--53.

\bibitem[{Livan, Novaes, and Vivo(2018)}]{Livan_2018}
Livan, G.; Novaes, M.; and Vivo, P. 2018.
\newblock \emph{Introduction to Random Matrices}.
\newblock Springer International Publishing.
\newblock ISBN 9783319708850.

\bibitem[{Lymburn et~al.(2021)Lymburn, Algar, Small, and Jüngling}]{lymburn21a}
Lymburn, T.; Algar, S.~D.; Small, M.; and Jüngling, T. 2021.
\newblock {Reservoir computing with swarms}.
\newblock \emph{Chaos: An Interdisciplinary Journal of Nonlinear Science}, 31(3): 033121.

\bibitem[{Mackey and Glass(1977)}]{mackey77a}
Mackey, M.~C.; and Glass, L. 1977.
\newblock Oscillation and Chaos in Physiological Control Systems.
\newblock \emph{Science}, 197(4300): 287--289.

\bibitem[{Mart{\'i}nez-Pe{\~{n}}a et~al.(2023)Mart{\'i}nez-Pe{\~{n}}a, Nokkala, Giorgi, Zambrini, and Soriano}]{pena23a}
Mart{\'i}nez-Pe{\~{n}}a, R.; Nokkala, J.; Giorgi, G.~L.; Zambrini, R.; and Soriano, M.~C. 2023.
\newblock Information Processing Capacity of Spin-Based Quantum Reservoir Computing Systems.
\newblock \emph{Cognitive Computation}, 15(5): 1440--1451.

\bibitem[{Mujal(2022)}]{mujal22a}
Mujal, P. 2022.
\newblock Quantum Reservoir Computing for Speckle Disorder Potentials.
\newblock \emph{Condensed Matter}, 7(1).

\bibitem[{Mujal et~al.(2023)Mujal, Mart{\'i}nez-Pe{\~{n}}a, Giorgi, Soriano, and Zambrini}]{mujal23a}
Mujal, P.; Mart{\'i}nez-Pe{\~{n}}a, R.; Giorgi, G.~L.; Soriano, M.~C.; and Zambrini, R. 2023.
\newblock Time-series quantum reservoir computing with weak and projective measurements.
\newblock \emph{npj Quantum Information}, 9(1): 16.

\bibitem[{Mujal et~al.(2021)Mujal, Martínez-Peña, Nokkala, García-Beni, Giorgi, Soriano, and Zambrini}]{mujal21a}
Mujal, P.; Martínez-Peña, R.; Nokkala, J.; García-Beni, J.; Giorgi, G.~L.; Soriano, M.~C.; and Zambrini, R. 2021.
\newblock Opportunities in Quantum Reservoir Computing and Extreme Learning Machines.
\newblock \emph{Advanced Quantum Technologies}, 4(8): 2100027.

\bibitem[{Nakajima, Tanaka, and Hashimoto(2021)}]{nakajima21a}
Nakajima, M.; Tanaka, K.; and Hashimoto, T. 2021.
\newblock Scalable reservoir computing on coherent linear photonic processor.
\newblock \emph{Communications Physics}, 4(1): 20.

\bibitem[{Runge(1895)}]{runge95a}
Runge, C. 1895.
\newblock Ueber die numerische Aufl{\"o}sung von Differentialgleichungen.
\newblock \emph{Mathematische Annalen}, 46(2): 167--178.

\bibitem[{Spagnolo et~al.(2022)Spagnolo, Morris, Piacentini, Antesberger, Massa, Crespi, Ceccarelli, Osellame, and Walther}]{spagnolo22a}
Spagnolo, M.; Morris, J.; Piacentini, S.; Antesberger, M.; Massa, F.; Crespi, A.; Ceccarelli, F.; Osellame, R.; and Walther, P. 2022.
\newblock Experimental photonic quantum memristor.
\newblock \emph{Nature Photonics}, 16(4): 318--323.

\bibitem[{Suzuki et~al.(2022)Suzuki, Gao, Pradel, Yasuoka, and Yamamoto}]{suzuki22a}
Suzuki, Y.; Gao, Q.; Pradel, K.~C.; Yasuoka, K.; and Yamamoto, N. 2022.
\newblock Natural quantum reservoir computing for temporal information processing.
\newblock \emph{Scientific Reports}, 12(1): 1353.

\bibitem[{Tanaka et~al.(2019)Tanaka, Yamane, Héroux, Nakane, Kanazawa, Takeda, Numata, Nakano, and Hirose}]{tanaka19a}
Tanaka, G.; Yamane, T.; Héroux, J.~B.; Nakane, R.; Kanazawa, N.; Takeda, S.; Numata, H.; Nakano, D.; and Hirose, A. 2019.
\newblock Recent advances in physical reservoir computing: A review.
\newblock \emph{Neural Networks}, 115: 100--123.

\bibitem[{Wang and Cichos(2024)}]{wang24a}
Wang, X.; and Cichos, F. 2024.
\newblock Harnessing synthetic active particles for physical reservoir computing.
\newblock \emph{Nature Communications}, 15(1): 774.

\end{thebibliography}

\end{document}